\documentclass[prb,aps,twocolumn,showpacs,superscriptaddress,floatfix]{revtex4}
\usepackage{eurosym}
\pdfoutput=1
\usepackage{epsfig}
\usepackage{amsmath}
\usepackage{amssymb}
\usepackage{amsfonts}
\usepackage{mathptmx}
\usepackage{eucal}
\usepackage{bm}
\usepackage{epstopdf}

\begin{document}

\title{Superconducting Phase Domains for Memory Applications}
\author{S.~V.~Bakurskiy}
\affiliation{Skobeltsyn Institute of Nuclear Physics, Lomonosov Moscow State University,
Leninskie gory, Moscow 119991, Russian Federation}
\affiliation{Faculty of Physics, Lomonosov Moscow State University, Leninskie gory,
Moscow 119992, Russian Federation}
\affiliation{Moscow Institute of Physics and Technology, Dolgoprudny, Moscow Region,
141700, Russian Federation}
\author{N.~V.~Klenov}
\affiliation{Skobeltsyn Institute of Nuclear Physics, Lomonosov Moscow State University,
Leninskie gory, Moscow 119991, Russian Federation}
\affiliation{Faculty of Physics, Lomonosov Moscow State University, Leninskie gory,
Moscow 119992, Russian Federation}
\affiliation{Moscow Institute of Physics and Technology, Dolgoprudny, Moscow Region,
141700, Russian Federation}
\author{I.~I.~Soloviev}
\affiliation{Skobeltsyn Institute of Nuclear Physics, Lomonosov Moscow State University,
Leninskie gory, Moscow 119991, Russian Federation}
\affiliation{Moscow Institute of Physics and Technology, Dolgoprudny, Moscow Region,
141700, Russian Federation}
\author{M.~Yu.~Kupriyanov}
\affiliation{Skobeltsyn Institute of Nuclear Physics, Lomonosov Moscow State University,
Leninskie gory, Moscow 119991, Russian Federation}
\affiliation{Moscow Institute of Physics and Technology, Dolgoprudny, Moscow Region,
141700, Russian Federation}
\author{A.~A.~Golubov}
\affiliation{Moscow Institute of Physics and Technology, Dolgoprudny, Moscow Region,
141700, Russian Federation}
\affiliation{Faculty of Science and Technology and MESA+ Institute for Nanotechnology,
University of Twente, 7500 AE Enschede, The Netherlands}
\date{\today }

\begin{abstract}
In this work we study theoretically the properties of S-F/N-sIS type
Josephson junctions in the frame of the quasiclassical Usadel formalism. The
structure consists of two superconducting electrodes (S), a tunnel barrier
(I), a combined normal metal/ferromagnet (N/F) interlayer and a thin
superconducting film (s). We demonstrate the breakdown of a spatial
uniformity of the superconducting order in the s-film and its decomposition
into domains with a phase shift $\pi $ . The effect is sensitive to the
thickness of the s layer and the widths of the F and N films in the
direction along the sIS interface. We predict the existence of a regime
where the structure has two energy minima and can be switched between them
by an electric current injected laterally into the structure. The state of
the system can be non-destructively read by an electric current flowing
across the junction.
\end{abstract}

\maketitle

Josephson junctions containing normal (N) and ferromagnetic (F) materials in
a weak link region are currently the subject of intense research. An
interest in such structures is due to the possibility of their use as
control elements of superconductor memory compatible with Single Flux
Quantum (SFQ) logic. A number of implementations of Josephson control
elements were proposed recently, among which the structures containing F
layers in the weak link region are of greatest interest \cite{Blamire,
Eschrig, LinderRev}. Various types of superconducting spin-valve structures
including two or more ferromagnetic layers have been proposed \cite%
{Tagirov,Gu,Bell,Oh,Qader, Baek, Robinson1, Krasnov, Alidoust,
Birge,Li,Gu1,FominovY,Zdravkov,Leksin}. The mutual orientations, parallel or
antiparallel, of magnetizations of the layers determine critical currents
and critical temperatures of the structures. Recently, it was proposed to
apply the phenomenon of triplet superconductivity in spin-valve devices with
noncollinear magnetization of the layers \cite{VolkovAF, Houzet,
Karminskaya, Gabor, Linder1,Khaire,Anwar,Blamire1,Klose}. The problem of
small characteristic voltage $I_{C}R_{N}$ in these structures was solved by
using an additional tunnel barrier connected through a thin superconducting
spacer \cite{Larkin,Bakurskiy1,Bakurskiy2,Ruppelt}. However, to control an
operation of these devices, the application of magnetic fields or strong
spin-polarized currents is necessary in order to switch the structure to a
different state. Such control requires the use of additional external
circuits resulting in restriction of possible memory density. Moreover,
characteristic operational times of such devices are limited by relatively
slow processes of the remagnetization of the ferromagnetic layers.

\begin{figure}[tb]
\begin{center}
\includegraphics[width=0.6\linewidth]{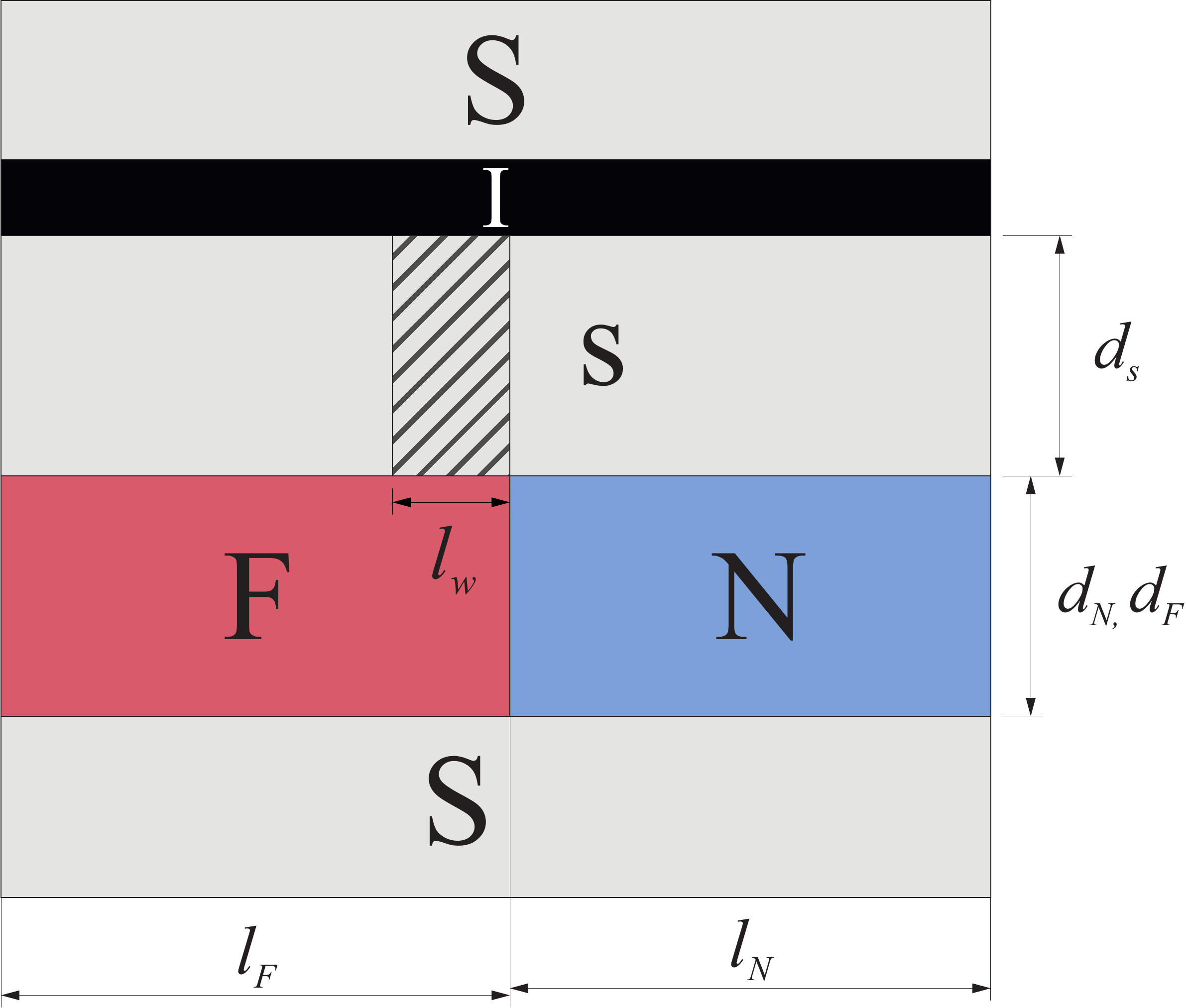}
\end{center}
\vspace{-3 mm}
\caption{The sketch of S-N/F-s-I-S structure. Thickness of the
superconducting, ferromagnetic and normal metal films are denoted as $d_s$, $%
d_N$ and $d_F$, respectively. The longitudional lengths of ferromagnetic and
normal metal films are denoted as $l_F$ and $l_N$. A probable area of SPD-wall formation is hatched in the figure and determined by the lentgth $l_w$. }
\vspace{-7 mm}
\label{picture}
\end{figure}

In this work we propose a S-N/F-s-I-S control unit for a superconducting
memory cell (Fig.\ref{picture}) based on the principles completely different
as the suggested earlier. The considered structure consists of the two
superconductive electrodes (S) and of the two weak link regions: the tunnel
barrier (I) and the metallic (N/F) interlayer. The (N/F) part is formed by
the longitudinally oriented normal (N) and ferromagnetic (F) layers. The
weak link areas are separated by a thin superconducting s-layer (see Fig.\ref%
{picture}). We consider the phenomenon of nucleation of the Superconducting
Phase Domains (SPD) in the thin s-layer induced by proximity effect with
the bulk superconducting S-electrode through the complex ferromagnetic
interlayer. We propose the way to control a switching between the SPD-state and
the single domain state of the junction. To provide the quantitative model of the
structure, we solve self-consistently the two-dimensional boundary-value
problem in the frame of the quasiclassical Usadel equations in the diffusive
regime \cite{Usadel, RevG, RevB, RevV, KL}.

\begin{figure}[tb]
\begin{center}
\includegraphics[width=0.9\linewidth]{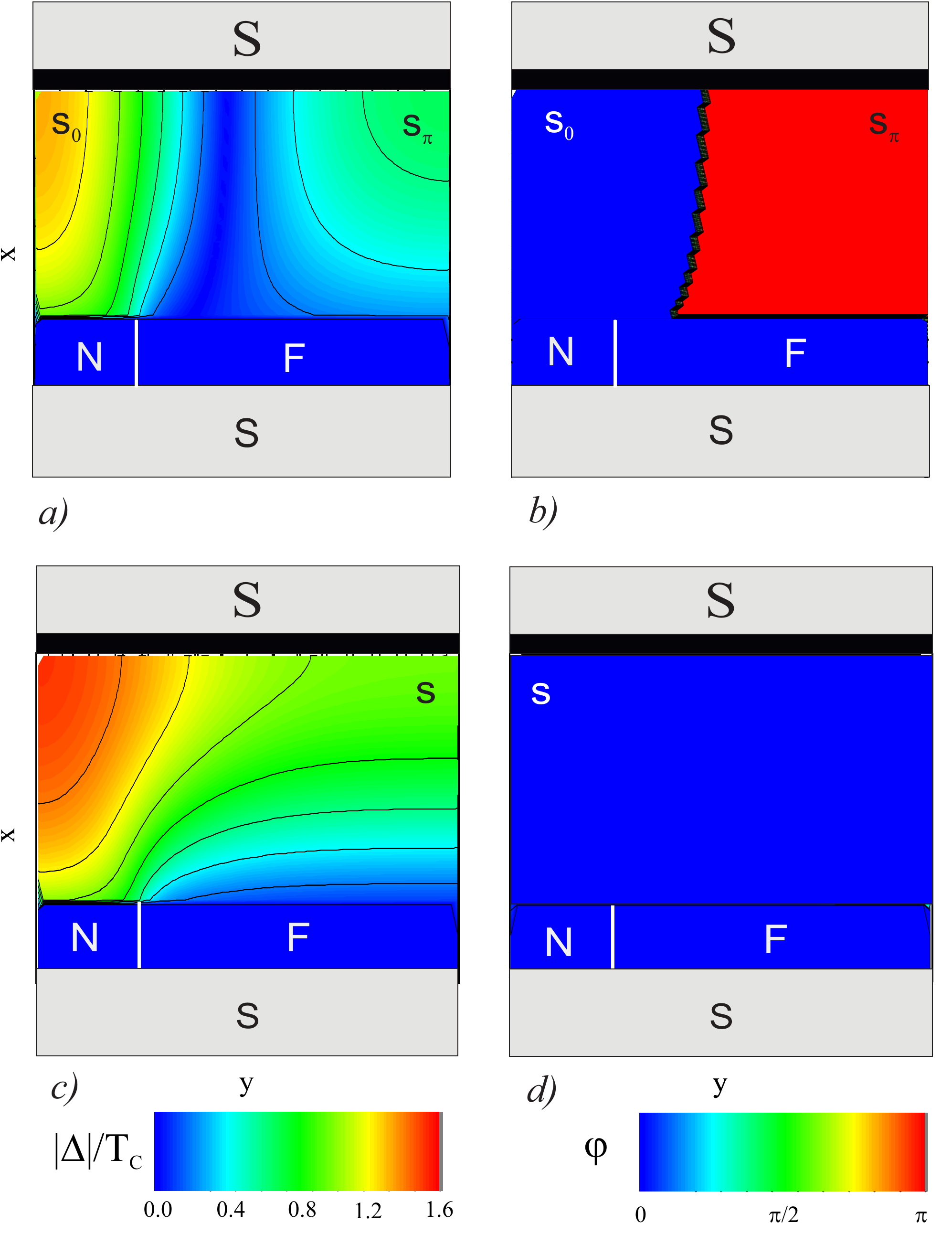}
\end{center}
\par
\vspace{-7 mm}
\caption{The distribution of the magnitude (a, c) and the phase (b, d) of the pair potential $\Delta$ over the S-F/N-s-I-S structure for various thicknesses of s-layer: (a, b) for $d_s=3.5\protect%
\xi_S$ and (c, d) for $d_s=4\protect \xi_S$. Thinner
superconducting layers are separated into two domains of $0$ and $\protect\pi$
phases, while thicker films stay in the single domain state with the $0$ phase.
The chosen parameters of the structure are: $d_F = 1\protect\xi_F, l_F = 12%
\protect\xi_S, l_N = 4 \protect\xi_S$, the exchange energy $H=10 \protect\pi T_C$
and the suppression parameter\cite{KL} $\protect\gamma_B=0.3$. }
\vspace{-7 mm}
\label{Delta2D}
\end{figure}

\textbf{Superconducting Phase Domains.} Let's consider the S-N/F-s-I-S
structure shown in Fig.\ref{picture}. We suppose that dirty limit conditions
are fulfilled for all metals and use the Usadel equations \cite{Usadel} with
Kupriyanov-Lukichev \cite{KL} boundary conditions at the interfaces. To
simplify the calculations, we also assume that all metals have equal
resistivities $\rho _{F}=\rho _{N}=\rho _{S}$ and the coherence lengths $\xi
_{F}=\xi _{N}=\xi _{S}=\sqrt{D_{S,N,F}/\pi T_C}$, while the ferromagnetic material is characterized
by the exchange energy $H$. In addition, we neglect suppression of
superconductivity in the S electrodes due to the inverse proximity effect,
while the superconducting properties of the middle s-layer are found in a
self-consistent manner. Finally, we assume that all spatial dimensions of
the structure are much smaller than the Josephson penetration depth $\lambda
_{J}$ and there are no vortices inside the structure.

The state of the S-N/F-s-I-S structure significantly depends on thickness of
the intermediate s-layer. In Fig.\ref{Delta2D} the results of numerical
calculations of the pair potential in the intermediate N/F-s layer are
demonstrated for different thicknesses of the s-layer. On the bottom panels
(c) and (d) the spatial distributions of the pair potential in the
relatively thick s-film with $d_{s}=4\xi_S$ are shown. The calculations
demonstrate that only the magnitude of the s-film order parameter varies in
space, while there is no spatial variation of its phase $\varphi $. This
state is defined as the $0$-state of the junction.

For thinner s-film with $d_{s}=3.5\xi_S$, two superconducting domains
separated by the area of suppressed superconductivity are clearly
distinguished (Fig.\ref{Delta2D}a, b). The superconducting phase $\varphi $
varies between these domains: $\varphi =0$ in the domain in the vicinity of
the N-layer and $\varphi =\pi $ in the vicinity of the F-layer. We call this
state as the Superconducting Phase Domain State (SPD-state) and define the domains
as $s_{0}$ and $s_{\pi }$, respectively.

The nature of this phenomenon can be understood in terms of the SFs and SNs
junctions connected in parallel. It is well known that the $\pi $-state
might be realized in the SFs part of the structure \cite{Ryazanov}. In the
case of a thick s-layer, we deal with the $0$-$\pi $ junction with the SFs
and SNs channels connected in parallel. Such configurations have been
discussed earlier in relation to possible realization of $\varphi $%
-junctions \cite{Buzdin,Gold,Bakurskiy3,Gold2,Soloviev}. The ground state
phase $\varphi $ of the junction depends on the relation between the
Josephson energies of the SFs and the SNs parts
\begin{equation}
E_{JF}=\frac{\hbar j_{CF}l_{F}W}{2e},\;E_{JN}=\frac{\hbar j_{CN}l_{N}W}{2e},
\label{E1}
\end{equation}%
where $j_{CF,N}$ are the critical current densities of the SFs and SNs
junctions, respectively, and $l_{F,N}W$ are their cross-section areas. Since
the energy of the SNs-junction is typically larger than that of the SFs one,
the SNs-junction controls the phase $\varphi =0$ of the s-layer. As long as we
consider the regime when the SFs-junction is in a $\pi $-state, the phase $%
\varphi =0$ is energetically unfavorable for the SFs part of the device.

The situation may drastically change with the decrease of the s-layer thickness.
As will be shown below, at a certain critical thickness the
s-film splits into two separate superconducting domains with phases shifted
by $\pi $.

To perform simple estimates of the critical s-layer thickness
and the thickness of the domain wall, one should compare the energies of SFs junction $E_{JF}$
and of the domain wall $E_{SPD}$. The latter has two contributions
\begin{equation}
E_{SPD}=E_{w}+2E_{Jw}=\Delta F_{w}l_{w}d_{s}W+\frac{\hbar j_{Cw}d_{s}W}{e}.
\label{E2}
\end{equation}%
The first one, $E_{w}$, is the superconducting condensation energy of the
volume with suppressed superconductivity. This term is proportional to the
volume of the domain wall and increases linearly with the growth of its thickness $%
l_{w}$. The second term $E_{Jw}$ is the Josephson coupling energy of the
junction between $0$ and $\pi$ domains. This term
decreases with the growth of $l_{w}$ due to the
suppression of the Josephson current density $j_{Cw}\sim Exp(-l_{w})$.
Minimizing the $E_{SPD}(l_{w})$ functional, one can estimate the width $l_{w}$,
which is of the order of $\xi _{S}$, in agreement with the results of the numerical
calculations (Fig.\ref{Delta2D}a). Further, using the condition of SPD-state formation
\begin{equation}
E_{w}+2E_{Jw}\leq 2E_{JF}  \label{conSPD}
\end{equation}%
and taking the typical junction parameters from Fig.\ref{Delta2D}, one can
estimate characteristic value of the critical s-layer thickness, which is of
the order of few $\xi _{S}$.

One has to take into account that in same range of s-layer thicknesses superconductivity in the s-layer
might be suppressed due to an inverse proximity effect \cite{Bakurskiy1,Bakurskiy2}.
However, since the critical thickness $d_{s}$ of the SPD-state formation is proportional to the lateral size $l_F$,
the latter can be chosen large enough to tune $d_{s}$ to larger values and thus to avoid suppression of superconductivity.
Numerical calculations show that $l_F=12 \xi _{F}$ is sufficient to observe the SPD state.

Further, we note that since the lateral size of the junction is smaller than $\lambda_J$,
one can neglect magnetic fields generated by currents in the structure and therefore disregard half-quantum vortices
which might be trapped in the junction and destroy the SPD state.

\textbf{Control unit for memory element. }In this section we discuss
possible application of the S-N/F-s-I-S device as a control unit for a
memory element. While the thickness of the intermediate s-layer $d_{s}$ is
close to the critical one, the S-F/N-s-I-S system stays in the vicinity of
the transition between $0$- and SPD-states.

\begin{figure}[t]
\begin{center}
\includegraphics[width=0.8\linewidth]{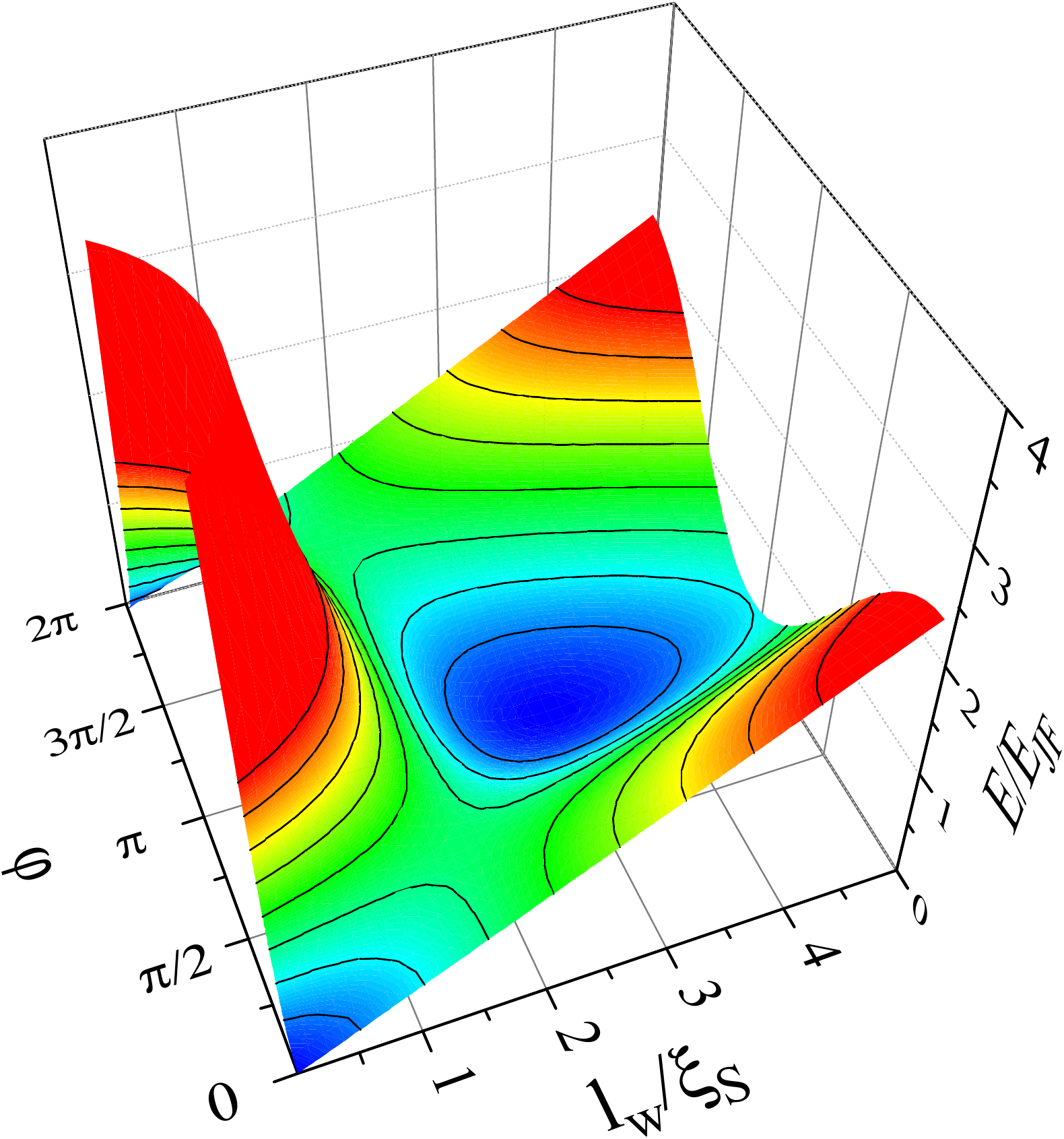}
\end{center}
\vspace{-4 mm}
\caption{Energy profile in the S-F/N-s-I-S structure versus phase $\protect\varphi$
of the $s_{\protect\pi}$ domain and thickness of the domain wall $l_{w}$. The parameters of the SPD-state $E_{Jw}=0.25 E_{JF}$ and $%
E_w = 1.5 E_{JF} $ are chosen so that system is in the critical regime
between the $0$ and SPD-states. In this case, the double-well potential forms two
local minima: the minimum in the left corner corresponds to the $0$-state and the one in the middle
corresponds to SPD-state.}
\vspace{-7 mm}
\label{EnergyGraph}
\end{figure}

To model the system, we take into account the Josephson energies of the
junctions and the pairing energy of s-layer and we treat the domain wall as
a normal layer with finite thickness $l_{w}$. The energy of the system
depends on the thickness $l_{w}$ and on the phase $\varphi $ of the $s_{\pi
} $ domain, neglecting small Josephson energy of the sIS junction:
\begin{equation}
E= E_{JF} (1-\cos (\pi -\varphi )) + E_w + E_{Jw} (1-\cos \varphi ) -2
E_{JF}.  \label{E3}
\end{equation}
Fig. \ref{EnergyGraph} shows the energy profile of the system in the regime
when the thickness $d_{s}$ is close to its critical value calculated from
Eqs.(\ref{E2},\ref{E3}). It is seen that there are two local minima at the
phases $\varphi =0$ and $\varphi =\pi $, which correspond to the $0$- and SPD- states. The 0-state local minimum is obtained at $l_{w}=0$ and
corresponds to the absence of the region of suppressed superconductivity.
Contrary, the SPD- state is realized at a finite thickness $l_{w}$. These
states are separated by a potential barrier, which provides the possibility
to store an information.

\begin{figure}[b]
\begin{center}
\includegraphics[width=0.9\linewidth]{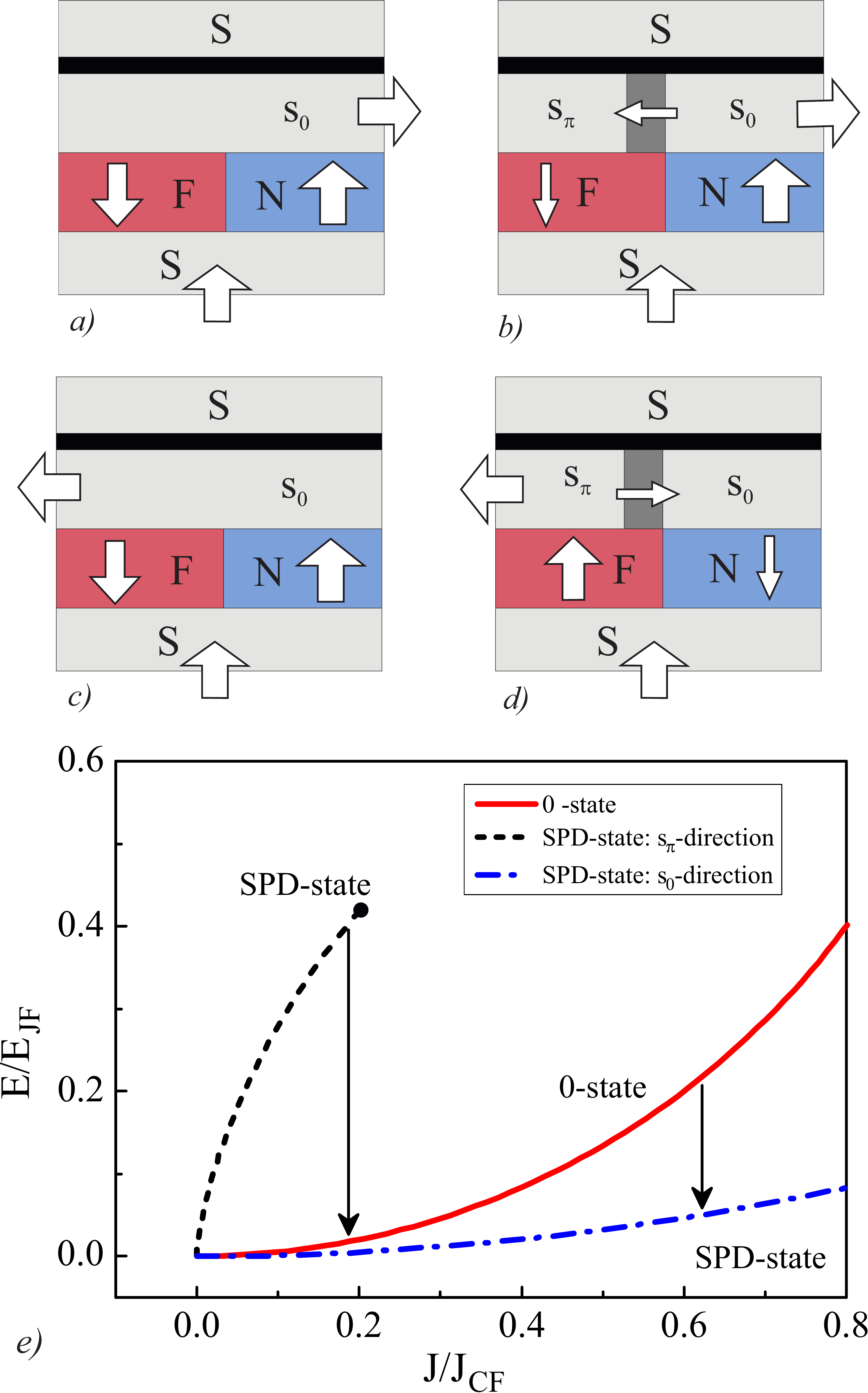}
\end{center}
\vspace{-5 mm}
\caption{ Schematic current distribution over the structure for the "Write"
operation in the $0$-state (a,c) and the SPD-state (b,d). The chosen parameters are $E_{Jw}=0.8 E_{JF}$, $E_{JN}=4 E_{JF}$ and $%
E_w = 0.4 E_{JF} $ The arrows demonstrate
the directions in which the current flows. e) Energies of the SPD and $0$- states versus
current for different connections of electrodes. Current $J$ is normalized on the critical current of the SFs part of the junction $J_{CF}$.
The solid line
corresponds to the 0-state (independent from type of connection), the dashed line
corresponds to the SPD state with electrode connected to $s_{\protect\pi}$
region, and dash-dotted line corresponds to the SPD- state with the electrode
connected to the $s_{0}$ domain.}
\label{Write}
\end{figure}

\textbf{\ "Write" and "Read" operations.} Below we propose how to perform
"Write" and "Read" operations in the considered device in the regime with
two local energy minima existing simultaneously (Fig. \ref{EnergyGraph}).
When an injected current is extracted through the right arm of the structure
(Fig. \ref{Write}a,b), the junction switches into the SPD-state.
Alternatively, for the left-side current extraction (Fig. \ref{Write}c,d),
the junction switches into the 0-state.

To illustrate the operation principle, we start from the 0-state (Fig. \ref%
{Write}a). Since the current phase relation of the SFs junction has an
additional $\pi$-shift, total Josephson current is the superposition of two
opposite supercurrents across the SNs and SFs parts of the junction. In the
competing SPD-state (Fig.4b) the total critical current is larger than in
the 0-state since the backflow through the SFs junction is limited by the
domain wall. As a result, switching from 0-state into the energetically
more favorable SPD-state should occur with an increase of the external
current, as illustrated in Fig.\ref{Write}e. When the current is switched off the
junction remains in the SPD-state.

In the case when the current is extracted from the left side, the
distribution of currents in the SPD-state significantly change (Fig. \ref%
{Write}d), and the junction behaves as a $\pi $ - junction with additional
backflow through the N-layer. In the competing 0-state Fig. \ref{Write}c,
the junction has higher value of the critical current and lower energy.
Therefore, switching from the SPD-state into the more favorable 0-state
should occur (see Fig.\ref{Write}e).

\begin{figure}[t]
\begin{center}
\includegraphics[width=0.8\linewidth]{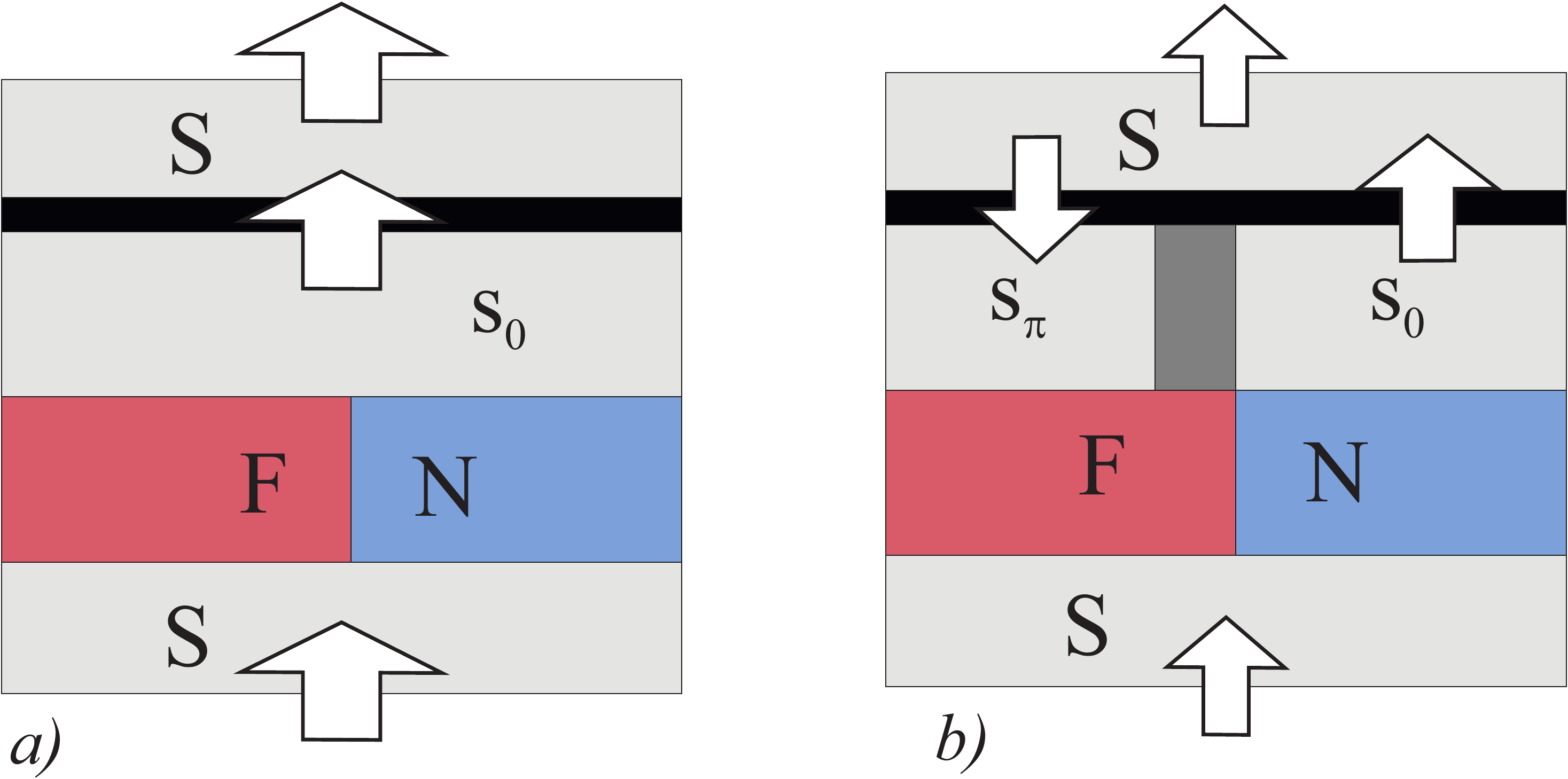}
\end{center}
\vspace{-5 mm}
\caption{ Schematic current distribution across the structure for the "Read"
operation in the $0$-state (a) and the SPD-state (b). The arrows demonstrate
directions in which the current flows. In the $0$-state the current is homogeneously
distributed over the tunnel barrier, while in the SPD-state the current is
separated over two channels with opposite directions of flow. Hence, the
latter state has the smaller critical current.}
\label{Read}
\vspace{-5 mm}
\end{figure}

A ''Read" operation can be implemented by a vertical current flowing across the
whole structure (Fig. \ref{Read}). In this case, the weak link of the
Josephson junction is located at the tunnel barrier I and the critical
current is much smaller than in the previously discussed cases. Hence, this
current can not change the state of the junction and the magnitude of the
critical current is determined only by the superconducting order parameters in
the vicinity of the tunnel barrier. In the $0$-state, the current is
distributed homogeneously over the whole tunnel barrier, while in the SPD%
-state, there are separated domains in the s-layer with a phase difference $%
\pi $ and the current through the tunnel barrier consists of two channels
with opposite current directions. The total current in the latter state is
much smaller than in the former. Therefore, the system can be used
as a control unit for a memory element.

\textbf{Discussion.} The performed analysis of the Josephson effect in
SIs(F/N)S structures suggests the possibility for the existence of
superconducting phase domains in the s-layer.

This phenomenon is not only of fundamental interest, but can also be used
for the realization of a control unit for superconducting memory cells. The
proposed unit has several noticeable advantages in comparison with the
existing solutions.

First, for its operation it is enough to have only one ferromagnetic film in
the weak link region. This opens the way for the realization of a
control unit with a large $I_{C}R_{N}$ product close to that for tunnel
junctions used in RSFQ logical circuits.

Second, switching between the equilibrium states doesn't require
remagnetization of the ferromagnetic layers, i.e. application of a strong
external magnetic field or spin-polarized currents. All "write" and "read"
operations are implemented by Josephson currents and never deal with time
scales specific to remagnetization processes. The characteristic time of the
considered device is based on mechanisms of destruction and recovery of
superconductivity in the thin s-film. This time is determined by properties
of electron-phonon interaction in the superconductor. These processes are
similar to those in superconducting single photon detectors\cite{Goltsman}
and can have a timescale in the order of $100-1000ps$ depending on material
constants of the s-layer.

Thus, the phenomenon of superconducting phase domains looks promising for
memory applications.

\textbf{Acknowledgment.} This work was supported by the Russian Science
Foundation, Project No. 15-12-30030.

\end{document}